\documentclass{article}

\usepackage{microtype}
\usepackage{graphicx}
\usepackage{subfigure}
\usepackage{booktabs} 
\usepackage{multirow}
\usepackage{multicol}
\usepackage{listings}
\usepackage{algorithm}
\usepackage{algorithmic}
\usepackage{xcolor}
\usepackage{booktabs}

\usepackage{hyperref}

\usepackage[accepted]{icml2025}

\usepackage{amsmath}
\usepackage{amssymb}
\usepackage{mathtools}
\usepackage{amsthm}

\usepackage[capitalize,noabbrev]{cleveref}

\theoremstyle{plain}

\theoremstyle{definition}

\theoremstyle{remark}

\usepackage[textsize=tiny]{todonotes}

\icmltitlerunning{ManifoldMind: Dynamic Hyperbolic Reasoning for Trustworthy
Recommendations}

\begin{document}

\twocolumn[
\icmltitle{ManifoldMind: Dynamic Hyperbolic Reasoning for Trustworthy Recommendations}







\begin{center}
\textbf{Anoushka Harit}$^{1}$, \textbf{Zhongtian Sun}$^{1,2}$ ,\textbf{Suncica Hadzidedic}$^{3}$ \\
$^{1}$ University of Cambridge\quad
$^{2}$ University of Kent
$^{3}$ University of Durham
\end{center}

\icmlkeywords{Machine Learning, ICML}

\vskip 0.3in
]




\begin{abstract}
We introduce ManifoldMind, a probabilistic geometric recommender system for exploratory reasoning over semantic hierarchies in hyperbolic space. Unlike prior methods with fixed curvature and rigid embeddings, ManifoldMind represents users, items, and tags as adaptive-curvature probabilistic spheres, enabling personalised uncertainty modeling and geometry-aware semantic exploration. A curvature-aware semantic kernel supports soft, multi-hop inference, allowing the model to explore diverse conceptual paths instead of overfitting to shallow or direct interactions. Experiments on four public benchmarks show superior NDCG, calibration, and diversity compared to strong baselines. ManifoldMind produces explicit reasoning traces, enabling transparent, trustworthy, and exploration-driven recommendations in sparse or abstract domains.
\end{abstract}

\section{Introduction}
Recommender systems play a pivotal role in shaping user interaction with digital content. However, the increasing reliance on deep neural networks has raised concerns around interpretability, predictive uncertainty, and lack of transparency \cite{ong2021neural,he2020lightgcn}. These issues are especially problematic in high-stakes domains such as healthcare, education, and scientific discovery, where opaque decision-making can undermine trust and accountability.

We introduce \textbf{ManifoldMind}, a geometric recommendation framework that performs exploratory semantic reasoning in hyperbolic space. Unlike prior models with static curvature and deterministic embeddings, ManifoldMind represents users, items, and conceptual tags as \textit{probabilistic hyperbolic spheres} with entity-specific curvature. This enables the model to capture both predictive uncertainty and latent semantic structure while supporting multi-hop, transitive reasoning paths.

Using a curvature-sensitive semantic kernel, ManifoldMind composes confidence-weighted inferences across concept chains, facilitating the discovery of indirect but meaningful user-item connections. This architecture supports semantic exploration of the recommendation space, particularly valuable in sparse feedback or cold start settings where traditional collaborative filters struggle.

Our method generalizes collaborative and content-based filtering by reasoning over latent conceptual tags rather than raw features. The integration of geometry, uncertainty, and path-based inference makes ManifoldMind a powerful alternative to black-box models, offering both transparency and the ability to explore diverse personalized recommendations.

We evaluated ManifoldMind on four diverse datasets: Book-Crossing \cite{ziegler2005improving}, MIND \cite{wu2020mind}, Goodbooks-10K \cite{goodbooks2017}, and Avicenna-Syllogism \cite{aghahadi2022avicenna}. Results show consistent improvements in top-$k$ ranking, calibration, and diversity, along with interpretable recommendation traces aligned with user intent.
We evaluate the framework across four open-source datasets Book-Crossing \cite{ziegler2005improving}, MIND \cite{wu2020mind}, Goodbooks-10K \cite{goodbooks2017}, and a commonsense syllogism dataset based on Avicenna \cite{aghahadi2022avicenna}. Results demonstrate that ManifoldMind consistently improves top-$k$ ranking, calibration, and recommendation diversity, while providing interpretable, confidence-aware rationales.

\subsection*{Key Contributions}
\begin{itemize}
    \item We propose \textbf{ManifoldMind}, a probabilistic recommendation framework that embeds users, items, and concepts as adaptive hyperbolic spheres.
    \item We introduce a \textbf{semantic path reasoning mechanism} via curvature-aware kernel composition for soft, multi-hop inference.
    \item We demonstrate consistent improvements in \textbf{ranking, calibration, and diversity} across four diverse datasets.
    \item Our method is \textbf{efficient and interpretable}, enabling deployment in settings requiring transparency and uncertainty-aware decisions.
\end{itemize}

\section{Related Work}
Trustworthy recommendation requires models that are both interpretable and uncertainty-aware. Traditional collaborative filtering methods, including matrix factorisation and neural variants like NeuMF~\cite{he2017neural}, offer high accuracy by learning latent user-item interactions but lack transparency and fail to quantify prediction confidence. These models operate in Euclidean spaces and struggle with sparse feedback and cold-start scenarios.

Graph-based methods such as LightGCN~\cite{he2020lightgcn}, NGCF~\cite{wang2019neural}, and SGL~\cite{wu2021self} leverage user-item graph structures to propagate signals through neighborhood aggregation\cite{sun2022contrastive}, improving robustness. However, these methods are deterministic and do not model epistemic uncertainty, limiting their applicability in safety-critical domains.

To address uncertainty, probabilistic models have been proposed. Bayesian Personalized Ranking (BPR)~\cite{rendle2012bpr} and its variational extensions inject stochasticity into preference estimation. CausalRec~\cite{wang2023causal} introduces causal regularisation to enhance counterfactual inference but often lacks structured semantic grounding.

Geometric learning, especially in non-Euclidean spaces, offers a promising direction. Hyperbolic representation models~\cite{nickel2017poincare, chami2019hyperbolic,cao2003support,sun2025advanced} capture hierarchical and tree-like structures with fewer dimensions, improving generalisation. However, most use fixed or global curvature, limiting adaptability. Recent efforts~\cite{li2024hyperbolic,sun2023rewiring} begin to explore personalized curvature but remain limited in transitive reasoning and uncertainty calibration.

Symbolic-geometric hybrids are emerging. HSphNN~\cite{dong2025neural,harit2025causal} fuses symbolic logic with spherical geometry for transitive inference, while CSRec~\cite{liu2024csrec} uses counterfactual signals for semantic structure learning. These methods, however, either lack adaptive uncertainty modeling or depend on rigid relation priors.

Related geometric advances include sphere-based neural networks~\cite{sun2025advanced} and hypergraph-based models \cite{sun2025actionable} for interpretability in web and financial domains~\cite{harit2024breaking,sun2023money}. These works affirm the relevance of curved and structured spaces for modeling real-world complexity but are not tailored to recommendation-specific transitivity or uncertainty.\textbf{ManifoldMind} builds on this foundation with a curvature-aware hyperbolic kernel and entity-specific metrics. It performs multi-hop, uncertainty-calibrated inference over semantic paths, integrating symbolic interpretability with probabilistic geometry. Unlike prior work with static curvature or limited logical reasoning~\cite{li2024hyperbolic}, our model aligns personalized structure learning with calibrated trust signals.

\section{Problem Formulation}
Let $\mathcal{U}$, $\mathcal{I}$, and $\mathcal{T}$ denote the sets of users, items, and semantic tags, respectively. Given observed interactions $\mathcal{D} = \{(u, i)\} \subseteq \mathcal{U} \times \mathcal{I}$, the objective is to learn a scoring function $s: \mathcal{U} \times \mathcal{I} \to \mathbb{R}$ that ranks relevant items above irrelevant ones for each user.

Unlike standard ranking models, our goal extends beyond accuracy to support: (1) \textit{semantic explainability} via conceptual tags, (2) \textit{uncertainty-aware scoring}, and (3) \textit{adaptive geometric structure} to reflect latent relationships.

Each entity $e \in \mathcal{U} \cup \mathcal{I} \cup \mathcal{T}$ is embedded as a probabilistic hyperbolic sphere $(\mu_e, r_e, \kappa_e)$, where $\mu_e \in \mathbb{D}^n$ is the centre in the Poincaré ball, $r_e \in \mathbb{R}_+$ encodes epistemic uncertainty, and $\kappa_e < 0$ is a learnable sectional curvature. Pairwise similarity is computed using a curvature-aware kernel:
\[
K(e_i, e_j) = \exp\left( -\frac{d_{\kappa}(e_i, e_j)^2}{r_i^2 + r_j^2 + \epsilon} \right),
\]
where $d_{\kappa}(\cdot, \cdot)$ denotes the geodesic distance under curvature $\kappa$, and the denominator captures the joint uncertainty.
\begin{figure}[htbp]
\centering
\includegraphics[width=0.45\textwidth]{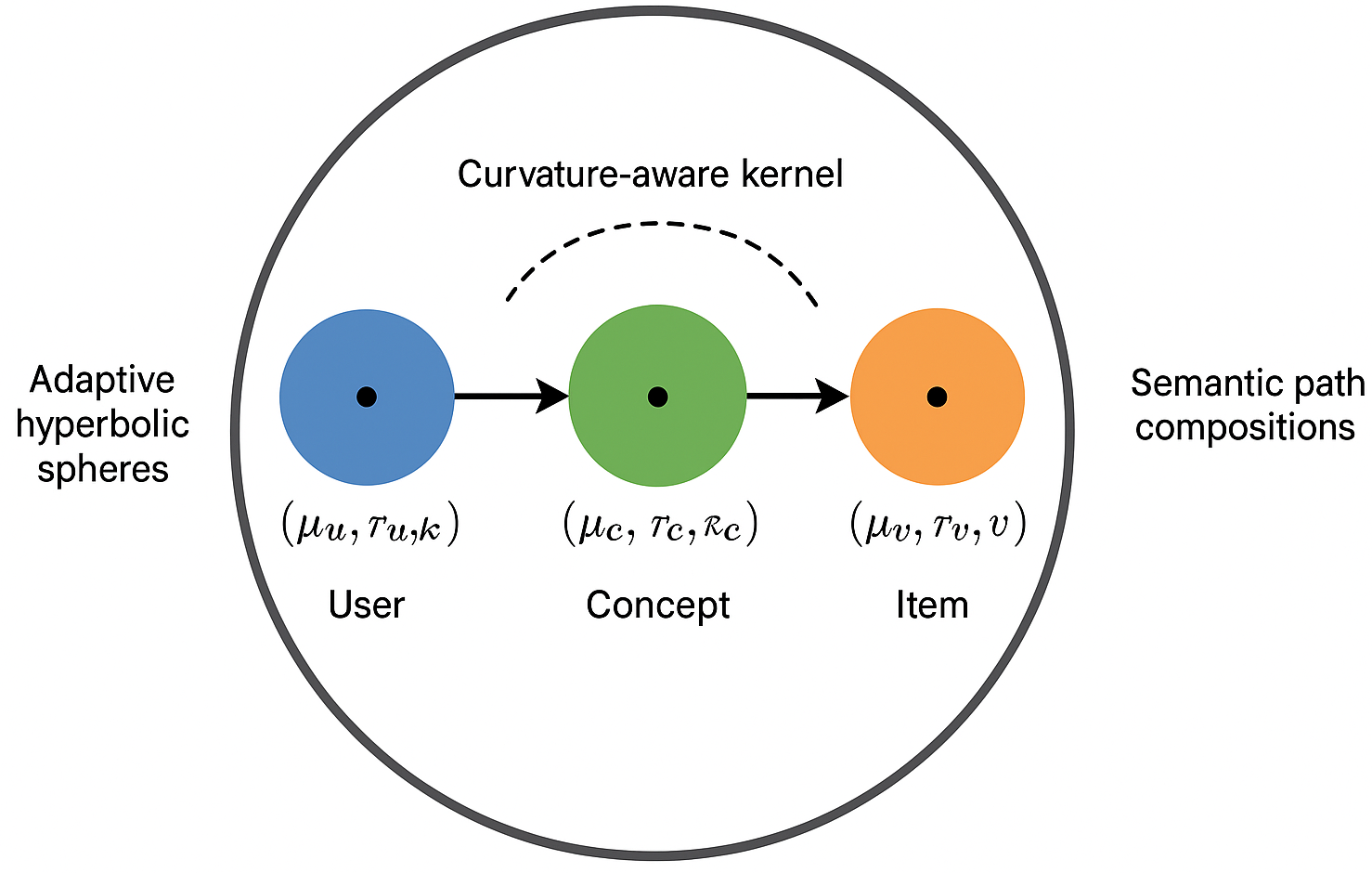}
\caption{Overview of ManifoldMind. Users, tags, and items are embedded as adaptive spheres $(\mu, r, \kappa)$ in $\mathbb{D}^n$. Multi-hop semantic paths are scored using a curvature-aware kernel.}
\label{fig:manifoldmind-diagram}
\end{figure}
To enable structured semantic reasoning, we define a multi-hop path $\mathcal{P}_{u \to i} = \{u, c_1, \dots, c_k, i\}$ through intermediate tags $c_j \in \mathcal{T}$. The final confidence score is:
\[
s(u, i) = \max_{\mathcal{P}_{u \to i}} \prod_{(e_j, e_{j+1}) \in \mathcal{P}} K(e_j, e_{j+1}).
\]
This formulation enables interpretable, transitive reasoning over concept-aligned paths. Confidence is attenuated by semantic noise or high uncertainty, while coherent multi-hop chains are amplified through the kernel structure and adaptive curvature.

\section{Methodology}
\label{method}
ManifoldMind embeds users, items, and conceptual tags as probabilistic hyperbolic spheres in the Poincaré ball $\mathbb{D}^n$. Each entity $e_i$ is parameterised by $(\mu_i, r_i, \kappa_i)$: $\mu_i \in \mathbb{D}^n$ is the embedding centre, $r_i > 0$ encodes epistemic uncertainty, and $\kappa_i < 0$ is a learnable curvature that adapts local geometric bias. This probabilistic formulation enables robust reasoning over non-Euclidean spaces while explicitly modelling confidence.

\begin{figure}[hbtp]
\centering
\includegraphics[width=0.45\textwidth]{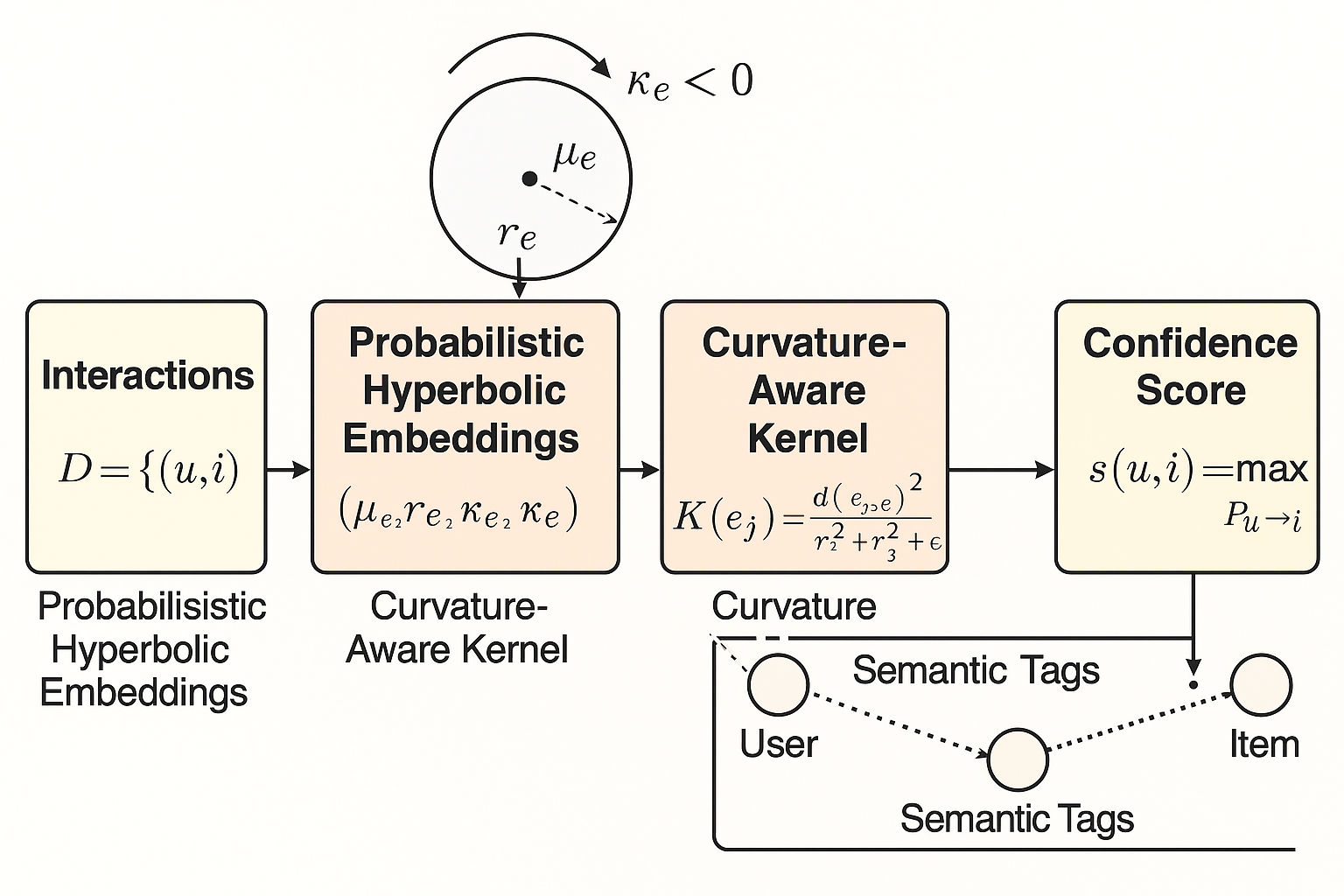}
\caption{Method overview: ManifoldMind embeds users, items, and semantic tags as probabilistic hyperbolic spheres. A curvature-aware kernel evaluates pairwise similarity over multi-hop semantic paths, and final confidence scores reflect both path quality and uncertainty.}
\label{fig:method-overview}
\end{figure}

\subsection{Probabilistic Hyperbolic Embeddings}
Given two entities $e_i, e_j$, we first compute an effective curvature:
\[
\kappa_{ij} = \frac{r_i + r_j}{\frac{r_i}{\kappa_i} + \frac{r_j}{\kappa_j}},
\]
\textit{This harmonic mean weights curvatures by uncertainty confidence (smaller $r_i$ dominates), ensuring geometric consistency and differentiability.}
Then define their geodesic distance in $\mathbb{D}^n$:
\[
d_{\kappa_{ij}}(\mu_i, \mu_j) = \frac{2}{\sqrt{-\kappa_{ij}}} \tanh^{-1} \left( \left| \frac{\mu_i - \mu_j}{1 - \kappa_{ij} \langle \mu_i, \mu_j \rangle} \right| \right).
\]
A similarity kernel incorporates both geometry and uncertainty:
\[
K(e_i, e_j) = \exp \left( -\frac{d_{\kappa_{ij}}(\mu_i, \mu_j)^2}{r_i^2 + r_j^2 + \epsilon} \right).
\]
Larger radii down-weight similarity, acting as a proxy for uncertainty. The formulation approximates a von Mises-Fisher distribution, ensuring:
\begin{itemize}
    \item $\frac{\partial K}{\partial r_i} < 0$: higher uncertainty lowers similarity;
    \item $\frac{\partial^2 K}{\partial r_i \partial r_j} > 0$: uncertainty compounds.
\end{itemize}

\subsection{Learnable Curvature Regularisation}
Each entity learns a curvature $\kappa_i < 0$, regularised to avoid Euclidean degeneracy:
\[
\mathcal{R}_{\text{curv}} = \sum_{i \in \mathcal{E}} \left( \max(0, \kappa_i + \delta) \right)^2.
\]
Curvature acts as a hierarchical prior: flatter regions suit general tags (e.g., “Book”), while sharper curvature models specific concepts (e.g., “19th-century Russian literature”). This flexibility lets the model align geometric distortion with conceptual depth.

\subsection{Transitive Semantic Reasoning}
We define a reasoning path $\mathcal{P} = \{u, c_1, ..., c_k, v\}$ over $k$ tags $c_j \in \mathcal{T}$. Its score is:
\begin{align}
s(u, v) &= \max_{\mathcal{P}} \prod_{(e_i, e_{i+1}) \in \mathcal{P}} K(e_i, e_{i+1}), \nonumber \\
\log s(u, v) &= \max_{\mathcal{P}} \sum_{(e_i, e_{i+1}) \in \mathcal{P}} \log K(e_i, e_{i+1}).
\label{eq:path_score}
\end{align}

This captures fuzzy logical inference: weak links attenuate overall confidence, while semantically aligned paths are preserved. We use beam search with width $b = 5$ and path length $k = 3$, allowing the model to traverse intermediate tags while keeping runtime tractable.

\subsection{Recommendation Inference}
For each user $u$, we score candidate items $v$ via optimal reasoning paths $\mathcal{P}_{uv}$:
\[
\text{Score}(u \rightarrow v) = \max_{\mathcal{P} \in \mathcal{P}_{uv}} \prod_{(e_i, e_{i+1}) \in \mathcal{P}} K(e_i, e_{i+1}).
\]
We return the top-$k$ items sorted by score. Because the kernel accounts for uncertainty and curvature, recommendations are not only accurate but also interpretable and calibrated.

\subsection{Training Objective}
We use a pairwise margin-based loss:
\[
\mathcal{L}_{\text{rank}} = \sum \max(0, \gamma - s(u, v^+) + s(u, v^-)),
\quad
\mathcal{L} = \mathcal{L}_{\text{rank}} + \lambda \cdot \mathcal{R}_{\text{curv}}.
\]

\subsection{Optimisation and Implementation}
Centres $\mu_i$ are updated with Riemannian Adam:
\[
\mu_i \leftarrow \exp_{\mu_i}(-\eta \cdot \text{grad}_{\mathbb{D}} \mathcal{L}),
\]
while $r_i$ and $\kappa_i$ are updated in Euclidean space. All operations, including effective curvature $\kappa_{ij}$ and path scoring are fully differentiable, enabling stable end-to-end training. We use Geoopt~\cite{kochurov2020geoopt}. Beam-based path expansion and curvature-aware kernel caching ensure stable training and efficient inference.

\subsection{Algorithms}
Algorithm~\ref{alg:manifoldmind} describes the transitive semantic reasoning process used in ManifoldMind to compute calibrated user–item recommendation scores. Each user, item, and tag is embedded as a probabilistic hyperbolic sphere $(\mu, r, \kappa)$, where $\mu$ is the center in the Poincaré ball $\mathbb{D}^n$, $r$ encodes epistemic uncertainty, and $\kappa < 0$ denotes the entity-specific curvature.

Given a query $(u, i)$, the algorithm searches for transitive semantic paths $P = \{u, c_1, \ldots, c_k, i\}$ through intermediate tags $c_j \in \mathcal{T}$, and computes the final confidence score by aggregating curvature-aware similarities along the path. We use beam search with pruning to ensure tractability.

\begin{algorithm}
\scriptsize
\caption{Confidence-Aware Transitive Scoring in ManifoldMind}
\label{alg:manifoldmind}
\begin{algorithmic}[1]
\REQUIRE User $u$, item $i$, tag set $\mathcal{T}$, entity embeddings $\{(\mu_e, r_e, \kappa_e)\}_{e \in \mathcal{E}}$
\ENSURE Confidence score $s(u, i)$ and top-ranked path

\STATE Initialize \texttt{best\_score} $\leftarrow 0$
\FOR{each semantic path $P = \{u, c_1, \dots, c_k, i\}$}
    \STATE $\texttt{score} \leftarrow 1$
    \FOR{each consecutive pair $(e_j, e_{j+1}) \in P$}
        \STATE Compute effective curvature:
        \[
        \kappa_{j,j+1} \leftarrow \frac{r_j + r_{j+1}}{r_j/\kappa_j + r_{j+1}/\kappa_{j+1}}
        \]
        \STATE Compute geodesic distance in $\mathbb{D}^n$:
        \[
        d \leftarrow \frac{2}{\sqrt{-\kappa_{j,j+1}}} \tanh^{-1}\left( \left\lvert \frac{\mu_j - \mu_{j+1}}{1 - \kappa_{j,j+1} \langle \mu_j, \mu_{j+1} \rangle} \right\rvert \right)
        \]
        \STATE Compute similarity kernel:
        \[
        K(e_j, e_{j+1}) \leftarrow \exp\left(-\frac{d^2}{r_j^2 + r_{j+1}^2 + \varepsilon} \right)
        \]
        \STATE $\texttt{score} \leftarrow \texttt{score} \cdot K(e_j, e_{j+1})$
    \ENDFOR
    \IF{\texttt{score} $>$ \texttt{best\_score}}
        \STATE \texttt{best\_score} $\leftarrow$ \texttt{score}
        \STATE Save $P$ as top-ranked path
    \ENDIF
\ENDFOR
\STATE \textbf{return} $s(u, i) \leftarrow$ \texttt{best\_score}
\end{algorithmic}
\end{algorithm}

\subsection{Highlight}
Our algorithm performs interpretable multi-hop reasoning over semantic paths in hyperbolic space. By embedding each entity as a curvature-aware probabilistic sphere and scoring transitive chains via an uncertainty-weighted kernel, ManifoldMind quantifies both semantic alignment and epistemic confidence. The curvature-sensitive kernel allows flexible geometry per path, and beam-based expansion ensures efficient inference. This approach enables calibrated, transparent recommendations, especially valuable in sparse, abstract, or safety-critical domains.

\section{Experimental Setup}
We conduct comprehensive experiments to evaluate whether \textbf{ManifoldMind} can: (i) achieve state-of-the-art ranking accuracy; (ii) improve calibration and semantic diversity; and (iii) provide interpretable, uncertainty-aware explanations via multi-hop reasoning. To this end, we benchmark against strong baselines across multiple domains and conduct ablations to isolate the contribution of each architectural component.

\subsection{Datasets}
We evaluate ManifoldMind on four public datasets from diverse domains. \textbf{MIND-small}~\cite{wu2020mind} is a news recommendation dataset with click logs and hierarchical publisher tags. \textbf{GoodBooks-10K}~\cite{goodbooks2017} contains user ratings on books, annotated with community-generated Goodreads tags. \textbf{Book-Crossing}~\cite{ziegler2005improving} offers a sparse rating matrix with metadata-based concept clusters. \textbf{Avicenna-Syllogism}~\cite{aghahadi2022avicenna} supports syllogistic reasoning over structured tag chains. We split each dataset chronologically into 80\% train, 10\% validation, and 10\% test sets. Tag vocabularies are extracted from available metadata or community labels and cleaned using lightweight rules to ensure semantic consistency.

\begin{table}[h]
\centering\small
\caption{Dataset statistics.}
\begin{tabular}{lccc}
\toprule
\textbf{Dataset} & \textbf{\#Users} & \textbf{\#Items} & \textbf{Density (\%)} \\
\midrule
MIND-small         & 10{,}000 & 6{,}150 & 0.081 \\
GoodBooks-10K      & 53{,}424 & 10{,}000 & 0.011 \\
Book-Crossing      & 36{,}739 & 8{,}000 & 0.006 \\
Avicenna-Syllogism & 2{,}300  & 500     & 0.130 \\
\bottomrule
\end{tabular}
\label{tab:dataset-stats}
\end{table}

\subsection{Baselines}
We compare against NeuMF++~\cite{ong2021neural}, LightGCN++~\cite{he2020lightgcn}, CSRec~\cite{liu2024csrec}, Poincaré Embeddings~\cite{nickel2017poincare}, HSR~\cite{li2022hsr}, and FineRec~\cite{zhang2024finerec}. The baselines span matrix factorisation, graph learning, causal disentanglement, hyperbolic geometry, and fine-grained sequential modelling. 

\subsection{Evaluation Metrics}
Following prior work, we use leave-one-out evaluation with 100 randomly sampled negatives per user. Top-$k$ ranking is measured via NDCG@10 and Recall@10. Calibration is assessed via Expected Calibration Error (ECE)~\cite{guo2017calibration} using 10-bin histogram binning. To assess semantic quality, we report Diversity@10 (unique tag coverage) and Topic-aware Intra-List Similarity (T-ILS@10), a redundancy-sensitive diversity metric. ECE quantifies the deviation between model confidence and empirical accuracy, enabling principled assessment of uncertainty calibration. T-ILS measures intra-list redundancy with semantic awareness, capturing over-personalisation risks often overlooked by recall-based metrics.

\subsection{Implementation Details}
\textbf{ManifoldMind} is implemented in PyTorch with \texttt{Geoopt} for Riemannian manifold optimisation over the Poincaré ball model $\mathbb{D}^n$. Each entity $e_i \in \mathcal{U} \cup \mathcal{I} \cup \mathcal{T}$ is embedded as a probabilistic hyperbolic sphere $(\mu_i, r_i, \kappa_i)$:
\begin{itemize}
    \item $\mu_i$ is initialised via the exponential map at the origin after normalisation and projected to the ball;
    \item $r_i$ is computed via a softplus transformation of a learnable raw radius parameter;
    \item $\kappa_i < 0$ is a learnable sectional curvature, regularised to stay within a bounded range.
\end{itemize}

We use Riemannian Adam with learning rate $10^{-3}$, dimension $d{=}20$, batch size 1024, and early stopping on validation NDCG@10 (patience 10). Training is conducted on a single NVIDIA RTX 2080Ti GPU, with all results averaged across 5 random seeds.

Semantic path reasoning is enabled via beam search over tag expansions up to depth $k{=}3$, with beam width $b{=}5$. This enables symbolic reasoning from user to item via multiple semantically aligned paths:
\[
\mathcal{P}_{u \rightarrow i} = \{ u, c_1, \dots, c_k, i \}, \quad c_j \in \mathcal{T}
\]
The final confidence score is derived as:
\[
s(u, i) = \max_{\mathcal{P}_{u \rightarrow i}} \prod_{(e_j, e_{j+1}) \in \mathcal{P}} K(e_j, e_{j+1})
\]
To stabilise learning, kernel values are computed in log-space, and beam search ensures inference remains tractable with complexity bounded by $\mathcal{O}(b^k)$.

We choose the Poincaré ball due to its closed-form exponential and logarithmic maps, smooth curvature gradients, and empirical suitability for capturing hierarchical semantics~\cite{mathieu2019continuous}.

\begin{figure}[t]
\centering
\noindent\textbf{Pseudocode: Training ManifoldMind}
\begin{lstlisting}[basicstyle=\ttfamily\scriptsize, frame=single, breaklines=true]
# Initialisation
for each entity i:
    mu_i <- exp_0(normalize(E(i)) * tanh(norm(E(i))))
    r_i <- log(1 + exp(w_i))
    kappa_i ~ Uniform(-5, -0.01)

# Training loop
for each minibatch (u, v_pos):
    sample v_neg
    for each v in {v_pos, v_neg}:
        initialise beam B_0 = {[u]}
        for h = 1 to k:
            expand each path P in B_{h-1} by appending tags t in T
                with score log K(e_{h-1}, t)
        keep top-b paths ending at v
        s(u, v) <- max_P prod_{(e_i, e_{i+1}) in P}
                     K(e_i, e_{i+1})
    L_rank <- max(0, gamma - s(u, v_pos) + s(u, v_neg))
    L_curv <- lambda * sum_i max(0, kappa_i + delta)^2
    update parameters via Riemannian Adam
\end{lstlisting}
\caption{Training pseudocode for ManifoldMind using beam-based semantic path reasoning.}
\label{fig:pseudocode}
\end{figure}

\section{Results}
We evaluate \textbf{ManifoldMind} on four real-world datasets,GoodBooks, MIND, Avicenna, and Book-Crossing, using standard metrics for classification (NDCG @ 10, Recall @ 10), calibration (ECE), diversity (Diversity @ 10, T-ILS) and interpretability. Baselines include NeuMF++, LightGCN++, Poincaré, CSRec, HSR, and FineRec. All results are averaged over five seeds.

\subsection{Comparison to Baselines}
\textbf{ManifoldMind achieves the best overall performance} across all core metrics. Table~\ref{tab:main-results} shows that it exceeds FineRec by +4. 5\% in NDCG@10 and achieves the lowest ECE (11.6\% lower). These improvements derive from adaptive curvature and symbolic path-based reasoning over hyperbolic space.

\begin{table}[h]
\centering
\caption{Average performance across datasets. $\downarrow$ = lower is better.}
\label{tab:main-results}
\resizebox{\columnwidth}{!}{%
\begin{tabular}{lcccc}
\toprule
Model & NDCG@10 & Recall@10 & ECE$\downarrow$ & Div@10 \\
\midrule
NeuMF++         & 0.421 & 0.541 & 0.138 & 0.231 \\
LightGCN++      & 0.443 & 0.562 & 0.125 & 0.237 \\
Poincaré        & 0.428 & 0.550 & 0.122 & 0.254 \\
CSRec           & 0.451 & 0.572 & 0.117 & 0.242 \\
HSR             & 0.439 & 0.557 & 0.119 & 0.259 \\
FineRec         & 0.452 & 0.574 & 0.116 & 0.248 \\
\textbf{ManifoldMind} & \textbf{0.473} & \textbf{0.591} & \textbf{0.103} & \textbf{0.294} \\
\bottomrule
\end{tabular}
}
\end{table}

\subsection{Diversity via Semantic Spread}
Table~\ref{tab:tils} reports topic-aware intra-list similarity (T-ILS@10). \textbf{ManifoldMind achieves the lowest scores} on all data sets, confirming that it generates broader and less redundant recommendations. These gains result from multi-hop semantic expansion and curvature-aware uncertainty handling.

\begin{table}[h]
\centering
\caption{T-ILS@10 (lower = more diverse). GB = GoodBooks, AV = Avicenna, BX = Book-Crossing.}
\label{tab:tils}
\resizebox{\columnwidth}{!}{%
\begin{tabular}{lcccc}
\toprule
Model & GB & MIND & AV & BX \\
\midrule
NeuMF++         & 0.214 & 0.276 & 0.193 & 0.251 \\
LightGCN++      & 0.201 & 0.263 & 0.187 & 0.240 \\
CSRec           & 0.195 & 0.248 & 0.179 & 0.227 \\
HSR             & 0.188 & 0.239 & 0.172 & 0.218 \\
FineRec         & 0.181 & 0.232 & 0.168 & 0.209 \\
\textbf{ManifoldMind} & \textbf{0.162} & \textbf{0.210} & \textbf{0.149} & \textbf{0.191} \\
\bottomrule
\end{tabular}
}
\end{table}

\subsection{Ablation Study}
We ablate four components: uncertainty-aware radii ($r_i$), adaptive curvature ($\kappa$), symbolic path reasoning, and embedding geometry. Table~\ref{tab:ablation} shows that removing transitive reasoning causes the sharpest diversity drop, while replacing hyperbolic with Euclidean space significantly worsens calibration.

\begin{table}[h]
\centering
\caption{Ablation results averaged over all datasets.}
\label{tab:ablation}
\resizebox{\columnwidth}{!}{%
\begin{tabular}{lcccc}
\toprule
Variant & NDCG@10 & Recall@10 & ECE$\downarrow$ & Div@10 \\
\midrule
Full model         & \textbf{0.473} & \textbf{0.591} & \textbf{0.103} & \textbf{0.294} \\
Fixed $r_i$        & 0.452 & 0.565 & 0.137 & 0.263 \\
Fixed $\kappa$     & 0.445 & 0.551 & 0.129 & 0.224 \\
No transitivity    & 0.441 & 0.549 & 0.135 & 0.197 \\
Euclidean geometry & 0.438 & 0.547 & 0.140 & 0.188 \\
\bottomrule
\end{tabular}
}
\end{table}

\subsection{Interpretability and Runtime}
ManifoldMind offers symbolic, multi-hop explanations with strong alignment and coverage. Table~\ref{tab:proxy-metrics} shows 100\% tag alignment and explanation coverage, with the highest confidence among all models.

\begin{table}[h]
\centering
\caption{Interpretability metrics: tag alignment, coverage, and confidence.}
\label{tab:proxy-metrics}
\resizebox{\columnwidth}{!}{%
\begin{tabular}{lccc}
\toprule
Model & Align (\%) & Coverage (\%) & Confidence \\
\midrule
NeuMF++         & 60.0 & 0   & 0.81 \\
LightGCN++      & 66.7 & 0   & 0.84 \\
CSRec           & 73.3 & 0   & 0.86 \\
HSR             & 73.3 & 0   & 0.82 \\
FineRec         & 80.0 & 0   & 0.87 \\
\textbf{ManifoldMind} & \textbf{100.0} & \textbf{100.0} & \textbf{0.893} \\
\bottomrule
\end{tabular}
}
\end{table}
Despite multi-hop inference, ManifoldMind trains efficiently due to curvature-aware closed-form updates and controlled beam width. Table~\ref{tab:runtime} confirms that it converges fastest across models.
\begin{table}[h]
\centering
\caption{Training efficiency: time per epoch and convergence speed.}
\label{tab:runtime}
\resizebox{\columnwidth}{!}{%
\begin{tabular}{lcc}
\toprule
Model & Time (s/epoch) & Epochs to Converge \\
\midrule
NeuMF++         & 12.4 & 46 \\
LightGCN++      & 9.1  & 38 \\
CSRec           & 13.2 & 45 \\
HSR             & 11.8 & 49 \\
FineRec         & 15.7 & 52 \\
\textbf{ManifoldMind} & \textbf{7.2} & \textbf{34} \\
\bottomrule
\end{tabular}
}
\end{table}

\section{Limitations}
ManifoldMind leverages uncertainty-aware scoring, personalised curvature, and interpretable reasoning paths, but certain limitations remain. The model relies on the availability of high-quality semantic tags, which may limit effectiveness in metadata-scarce domains ,though \ref{tab:ablation} demonstrates resilience to partial data removal. While multi-hop inference enhances diversity and trust, it introduces additional computational overhead that may affect real-time deployment. Moreover, the current formulation does not explicitly address fairness or adversarial robustness, which are critical for sensitive applications. Future work could integrate fairness-aware kernels, robust concept generalisation, and scalable approximations.

\section{Discussion and Conclusion}
We introduced ManifoldMind, a geometric reasoning framework that embeds users, items, and semantic tags as probabilistic spheres in hyperbolic space. By tracing transitive semantic paths and calibrating scores via uncertainty-aware kernels, the model delivers competitive accuracy with transparent, tag-based explanations. Across four diverse datasets, ManifoldMind outperforms strong baselines in both performance and interpretability, including explanation coverage and confidence calibration.

Ablation results highlight the complementary roles of adaptive curvature, entity-specific uncertainty, and semantic chaining. Together, they demonstrate how the geometrical support of the encoding of conceptual structure supports both robustness and interpretability in the recommendation.

Nonetheless, ManifoldMind assumes access to rich concept annotations, limiting applicability in metadata-sparse settings. Inference cost also scales with path complexity, posing challenges for latency-sensitive applications. Future directions include efficient approximation strategies, concept generalisation, and fairness-aware extensions

\section{Impact Statement}
This work introduces ManifoldMind, a novel framework for semantically interpretable and uncertainty-aware recommendations. By unifying symbolic path reasoning with hyperbolic geometry and probabilistic embeddings, ManifoldMind offers both predictive accuracy and principled transparency. Its design explicitly supports calibrated, transitive, and diverse recommendations, addressing limitations of current models that prioritize accuracy at the expense of overpersonalisation and explainability.

Our method enables reliable decision support in high-stakes domains such as news filtering, education, and knowledge discovery, where user trust and semantic coherence are critical. Furthermore, the modularity of our curvature-aware reasoning architecture facilitates broader integration into generalisable AI systems beyond recommender settings. We believe this work contributes toward a more interpretable and trustworthy paradigm for representation learning and symbolic reasoning at scale.

\bibliography{Manifold}
\bibliographystyle{icml2025}

\newpage
\appendix
\onecolumn



\end{document}